\documentclass[12pt]{article}
\usepackage{latexsym,graphicx,multirow}
\usepackage{amssymb}
\usepackage{amsmath}
\usepackage{amscd}
\usepackage{amsthm}
\usepackage{float}
\usepackage[left=2cm,top=2.5cm,right=2.5cm,bottom=1.5cm]{geometry}
\usepackage{graphicx}
\usepackage{epstopdf}
\usepackage{graphicx,epstopdf}
\usepackage{hyperref}
\DeclareGraphicsExtensions{.eps}
\usepackage{epstopdf} 
\usepackage{pdflscape}

\begin{document}

	\begin{center}
		\large{\bf
		Interacting Tsallis	holographic  dark energy: Cosmic behaviour, statefinder analysis and $\omega_{D}-\omega_{D}^{'}$ pair in the non- flat universe} \\
		\vspace{10mm}
		\normalsize{ Umesh Kumar Sharma$^1$,  Vipin Chandra Dubey$^2$, A. Pradhan$^3$  }\\
		\vspace{5mm}
		\normalsize{$^{1,2,3}$Department of Mathematics, Institute of Applied Sciences and Humanities, GLA University\\
			Mathura-281 406, Uttar Pradesh, India}\\
		
		\vspace{2mm}
		$^1$E-mail: sharma.umesh@gla.ac.in\\
         $^2$E-mail: vipin.dubey@gla.ac.in.\\
         $^3$E-mail: pradhan.anirudh@gmail.com\\
			\vspace{2mm}
		\vspace{5mm}
		\vspace{10mm}
	
\end{center}
\begin{abstract}
The paper explores the interacting Tsallis holographic dark energy (THDE) model in a non-flat universe following an infrared cut - off as the apparent horizon. The equation of state (EoS) and the deceleration parameter of THDE model are determined to understand the cosmological evolution for interacting THDE model in the nonflat universe. By applying the statefinder $(r, s)$ parameter-pairs diagnostic and $\omega_{D}-\omega^{'}_{D}$ pair dynamical analysis for the derived THDE model, we plot the evolutionary trajectories for different cases of Tsallis parameter $\delta$ and interaction term $b^{2}$ and also, for spatial curvature $\Omega_{k0}$= $ 0$, $-0.0012$ and $0.0026$ corresponding to flat, open and closed universes, respectively, in the framework  of Planck 2018  base cosmology results VI- LCDM  observational data.

 \smallskip 
 {\bf Keywords} : Interacting THDE, Statefinder, $\omega_{D}-\omega^{'}_{D}$ pair\\
 PACS: 98.80.Es, 95.36.+x, 98.80.Ck\\
\end{abstract} 
\section{Introduction}

In the standard cosmology scenario, the cosmological information from various observations and CMB experiments will be utilized to handle essential inquiries regarding the idea of the physical composition of the Cosmos \cite{ref1}. The sequence of cosmological observations for more than twenty years have shown that our current Universe is in accelerated expansion phase \cite{ref2,ref3,ref4,ref5,ref6}. At present, the fundamental scenario used to depict the accessible surveys is the CDM cosmology, a model where the dark matter, which around $25\%$ of the matter part of the universe and the dark energy (DE) \cite{ref6a}, approximately $70\%$ of the rest cosmic foundation, is responsible for the current accelerated stage of the Universe. This amazingly straightforward model can clarify the greater part of the various observations traversing over a vast area of length scales \cite{ref7}.\\

From the hypothetical perspective, notwithstanding, it is additionally outstanding that so as to give a decent depiction of the observed Universe the estimation of the energy density (vacuum), $\rho_{\Lambda} \equiv 10^{-47} $ Gev$^{4}$ prompts a disrupted circumstance in the interface among Particle Physics and Cosmology, and since it contrasts from hypothetical desires by 60-120 orders of magnitude \cite{ref8}. In addition, despite the fact that the advancement of these two dark segments over the astronomical time is essentially unique, their present energy densities are of a similar order, which leads to the inquiry whether this is just a fortuitous event or has an increasingly basic reason. Such inquiries are known as the cosmological constant problem \cite{ref9}. In this manner, given the hypothetical vulnerabilities on the nature and properties of the DE various components of astronomical acceleration have been examined, including alterations of gravity on enormous scales or then again a conceivable interaction between the parts of the dark sector.\\

Specifically, interacting models of dark matter and DE \cite{ref10,ref11,ref12,ref13,ref14} depend on the ground that no known symmetry in Nature anticipates or stifles a non-minimal coupling between these parts and, thusly, such probability must be explored in the framework of observational data (for a recent review, see \cite{ref15}). In certain classes of these coupled models, the coincidence issue previously mentioned may be generally eased when contrasted with the fundamental cosmology and firstly examined in \cite{ref12}, where the authors explored asymptotic attractor practices for the proportion of the DE and dark matter densities.
From that point forward, various interacting models with both analytical and numerical solutions have been suggested (see, for example \cite{ref10,ref11,ref13,ref16,ref17,ref18} and references therein).\\

Recently, a new model of dark energy called Tsallis holographic dark energy (THDE) \cite{ref19} has been proposed to explain the present accelerated expansion of the Universe using Tsallis generalized entropy, $ S_{\delta }=\gamma A^{\delta }$\cite{ref20}. The foundation of the HDE approach is the definition of the system boundary, and
in fact, modification in the HDE models can be done by changing the system entropy, the more details about HDE models can see the review \cite{ref21}. Cohen et al. \cite{ref22}, detailed a relation between the entropy S, the UV cutoff ($\Lambda$) and the IR cutoff $L$ is $L^{3} \Lambda^{3}\leq S^{\frac{3}{4}}$, which after combining with $S_{\delta}= \gamma A^{\delta}$, leads to $\Lambda^{4} \leq (\gamma(4\pi)^{\delta})L^{2\delta-4}$.
By the application of this inequality, energy density of THDE model leads
$\rho_{T} = CL^{2\delta-4}$ where C is a parameter which is unknown \cite{ref23,ref24,ref25}. Obviously, the THDE model has one more parameter $\delta$ than the
standard HDE model. THDE models with different IR cutoff have been investigated and explored in different scenario \cite{ref26,ref27,ref28,ref29,ref30,ref31,ref32,ref33,ref34,ref34a,ref34b,ref34c,ref34d}.\\

Naturally, how to distinguish the different kinds of DE models, their interaction effect and the
various model parameters in one model becomes an interesting subject. Moreover, the differences between the standard
$\Lambda$CDM model and other DE models are also attractive because today's observations are mostly based on the $\Lambda$CDM
model. Thus, the diagnostic methods for the DE models have been widely researched. The common methods are the
geometrical diagnostic tools called the statefinder diagnostic \cite{ref35,ref36} related to the third derivative of
the scale factor a(t). On the other hand, because of the characteristic of the EoS for the DE models,
$\omega_{D}-\omega_{D}^{'}$ analysis \cite{ref37} can also be used to distinguish various models. As explored in
\cite{ref38,ref39,ref40,ref41,ref42,ref43,ref44,ref45,ref46} the statefinder may effectively differentiate between
a wide variety of dark energy models including quintessence, the cosmological
constant, braneworld and the Chaplygin gas models and
interacting dark energy models. Recently, Sheykhi (2018), have derived the modified Friedmann equations for an FRW universe with the apparent horizon as IR cutoff in the form of Tsallis entropy \cite{ref47}. The geometric behaviour of non-interacting and
interacting THDE have diagnosed
for a flat universe in terms of statefinder parameters and
$\omega_{D} - \omega_{D}^{'}$ pair in detail \cite{ref48,ref49} considering IR cutoff as Hubble horizon. The effectiveness of the different diagnostic
methods for the three different cut-offs (the future event horizon, the Hubble horizon and the
GO horizon) of THDE models examined with different forms of interaction term $Q$ \cite {ref50}. The statefinder diagnosis and $\omega_{D} - \omega_{D}^{'}$ plane diagnostic to the non - interacting THDE model has been investigated
in the non-flat universe with the apparent horizon as IR cutoff \cite {ref51}.\\

Based on the above motivations, in this work, we apply the statefinder diagnosis and $\omega_{D} - \omega_{D}^{'}$ plane diagnostic to the interacting THDE model
in the nonflat Universe with the apparent horizon as IR cutoff. The statefinder may likewise be utilized to analyze distinctive illustration of the model, including different model parameters and distinctive spatial curvature ($\Omega_{k0}$) inputs.
Moreover, we consider the values of
$\Omega_{k0} =$ $0$, $-0.0012$ and $0.0026$
corresponding to the flat, open and closed universes, respectively, in the light of Planck 2018 results VI- LCDM base cosmology \cite{ref6}\\

The sequence of the paper is as follows: In Sec. 2, we review the
THDE considering the IR cutoff as the apparent horizon. In Sec. 3, observational results for spatial curvature is given. In Sec. 4, the cosmic behaviour of the deceleration parameter and EoS is described. In Sec. 5,
the statefinder parameters and $\omega_{D} - \omega_{D}^{'}$ plane are explored, plotted and discussed. Conclusions are talked about in Sec. 6\\

\section{  The THDE model with IR cutoff as apparent horizon}

The metric for FRW non-flat universe is defined as :
\begin{eqnarray}
\label{eq1}
ds^{2} = -dt^{2}+a^{2}(t)\Big(\frac{dr^{2}}{1-kr^{2}}+ + r^{2}d\Omega^{2}\Big),
\end{eqnarray}
where $ k = -1$, $ 1 $ and $0$ represent a open, closed and flat universes, respectively.
The first Friedmann equation in a non-flat FRW universe, including THDE and darkmatter (DM) is given as :

\begin{eqnarray}
\label{eq2}
H^2 + \frac{k}{a^2}=\frac{1}{\tilde{r}^2{}_A}=\frac{1}{3} (8 \pi G) \left(\rho_D+\rho _m\right),
\end{eqnarray}

where $\rho_{m}$ and $\rho_{D}$ represent the energy density of matter and THDE, respectively, and $\frac{\rho_{m}}{\rho_{D}} = r$ represent the ratio of energy densities of two dark sectors \cite{ref15,ref31}. Using the fractional energy densities, the energy density parameter of pressureless matter, THDE and curvature term can be expressed as \\

\hspace{2cm} $\Omega_{m} = \frac{8\pi\rho_{m}G}{3H^{2}} $, \hspace{1cm} $\Omega_{D} = \frac{8\pi\rho_{D}G}{3H^{2}} $, \hspace{1cm} $\Omega_{k} = \frac{k}{a^{2}H^{2}}$.\\

  Considering the apparent horizon as IR cutoff as the usual system boundary for the FRW universe situated at \cite{ref52,ref53,ref54}.
 \begin{eqnarray}
 \label{eq3}
 \tilde{r}_A=\frac{1}{\sqrt{\frac{k}{a^2}+H^2}}.
 \end{eqnarray}
Generalizing Boltzmann- Gibbs entropy to the non-additive entropy, called Tsallis entropy.  The black hole horizon entropy can be
 modified as \cite{ref20}
 \begin{eqnarray}
 \label{eq4}
 S_{\delta }=\gamma A^{\delta }
 \end{eqnarray}
 where $\delta$ represents the non-additive parameter and $\gamma $ denotes unknown constant. Since a null hypersurface is represented by $\tilde{r}_A$ for the FRW spacetime
 and also, a proper system boundary
 \cite{ref52,ref53,ref54}, a property equivalent to that of the black hole horizon, one may observe
 $\tilde{r}_A$ as the IR cutoff, and utilize the holographic DE speculation to get \cite{ref27,ref31,ref47}.
 \begin{eqnarray}
 \label{eq5}
 \rho _D=C\tilde{r}_A{}^{2 \delta -4},
 \end{eqnarray}
 where $C$ denotes the unknown parameter. It was contended that the entropy related with the
 apparent horizon of FRW universe, in every gravity hypothesis, has a similar structure as the same structure as the entropy of black hole horizon in the relating gravity. The
 just change one need is supplanting the black hole horizon span $ r_+ $ by the apparent
 horizon radius $\tilde{r}_A$ \cite{ref31}.\\

 Now Eq. (\ref{eq2}) can be written as:
 \begin{eqnarray}
 \label{eq6}
 1 +\Omega _k = \Omega _D+\Omega _m.
 \end{eqnarray}
 The law of conservation for the  interacting matter and THDE are given as :
 \begin{eqnarray}
 \label{eq7}
\dot \rho_{m} + 3 H \rho_{m} = Q,
 \end{eqnarray}
 \begin{eqnarray}
 \label{eq8}
 \dot \rho_{D} + 3H (\rho_{D} + p_{D}) = - Q.
 \end{eqnarray}
 It should be noted that the observation today
 also allows a mutual interaction $Q$ between DE and DM. Thus, $Q$ can be embedded in the THDE models.  We also consider \\
 
\hspace{4cm} $ Q = 3 b^{2} H(\rho_{D}+\rho_{m}) = 3 b^{2} H\rho_{D} (1 +r) $.\\
 
The ratio $ p _D/\rho _D= \omega _D $ represents the THDE EoS parameter. Combining with the definition of $r$, we get

 \begin{eqnarray}
 \label{eq9}
 r=\frac{\Omega _K+1}{\Omega _D}-1.
 \end{eqnarray}
 
 Now, using derivative with time of Eq. (\ref{eq2}) in Eq. (\ref{eq8}), and combined the result with
 Eqs. (\ref{eq7}) and  (\ref{eq6}), we get

 \begin{eqnarray}
 \label{eq10}
 \frac{\dot{H}}{H^2}=\Omega _K-\frac{3}{2} \Omega _D \left(\omega _D+r+1\right).
 \end{eqnarray}
 
 Using Eq. (\ref{eq10}), The deceleration parameter $q$ is obtained as
 \begin{eqnarray}
 \label{eq11}
 q=-\frac{\dot{H}}{H^2}-1=\frac{3}{2} \Omega _D \left(\omega _D+r+1\right)-\Omega _K-1.
 \end{eqnarray}
 Now, using the derivative with time of Eq. (\ref{eq5}) with Eqs. (\ref{eq3})
 and Eq. (\ref{eq10}), we get
 \begin{eqnarray}
 \label{eq12}
 \dot{\rho _D}=\frac{3 (\delta -2) H \rho _D \Omega _D \left(\omega _D+r+1\right)}{\Omega _K+1}.
 \end{eqnarray}
 Now by using the time derivative of energy density parameter $\Omega _D $ with Eqs. (\ref{eq12}) and (\ref{eq10}), one gets
 
 \begin{eqnarray}
 \label{eq14}
 \Omega _D'=\Omega _D \left(\frac{3 \Omega _D \left(\omega _D+r+1\right) \left(\delta +\Omega _K-1\right)}{\Omega _K+1}-2 \Omega _K\right)
 \end{eqnarray}
 where, dot in previous equation denotes derivative with respect to time, and prime denotes the derivative with respect to the ln a.\\
 
 Additionally, calculations for the density parameter and EoS parameter are given as
 
\begin{eqnarray}
\label{eq17}
\Omega '_{D}= \frac{\Omega _D \left(\Omega _D \left(-3 \delta +(1-2 \delta ) \Omega _K+3\right)-\left(\Omega _K+1\right) \left(3 \left(b^2-1\right) (\delta -1)+\left(3 b^2-1\right) \Omega _K\right)\right)}{(\delta -2) \Omega _D+\Omega _K+1}.
\end{eqnarray}

 \begin{eqnarray}
 \label{eq18}
{\scriptscriptstyle  \omega _D'= -\frac{3 (\delta -1) \left(\Omega _K+1\right) \left(\left(b^2-1\right) \left(\Omega _K+1\right)+\Omega _D\right) \left(2 b^2 (\delta -2) \Omega _D \left(\Omega _K+1\right)+b^2 \left(\Omega _K+1\right){}^2 +(\delta -2) (\delta -1) \Omega _D^2\right)}{\Omega _D \left((\delta -2) \Omega _D+\Omega _K   +1\right){}^3}}.
 \end{eqnarray}

\section{\bf Spatial curvature : Observational results}

There are additionally enough inspirations for considering a nonflat universe. Even though, it is generally accepted that curvature effect in the early phase of the universe is washed out practically by inflation. But, it does not really suggest that the curvature must be entirely ignored at present. In this segment, we present the observational information for spatial curvature utilized in the examination of THDE model.\\

By the simplest inflationary models, the spatial hypersurfaces are flat based on the $\Lambda$CDM base model assumption, predicted within measurable precision. This prediction may be examined with the combination of BAO and CMB
observational data to high accuracy. The Cosmic Microwave Background (CMB) alone experiences the geometric degeneracy, which is feebly shattered by the expansion of CMB lensing. The blend of polarization power spectra with the Planck temperature give\\

\hspace{3cm} $\Omega_{k} = -0.056^{+0.028}_{-0.018}$ \qquad ( Planck TT + lowE, 68 \%)\\

\hspace{3cm} $\Omega_{k} = -0.044^{+0.018}_{-0.015}$ \qquad ( Planck TT, TE, EE + lowE, 68 \%)\\

an apparent curvature detection at 2 $\sigma$ level. With only about $1/10000$ samples at $\Omega_{k} \geq 0$, for the $TT, TE, EE + lowE$ result is $-0.095 < \Omega_{k} < -0.007$ at the $99\%$
probability region \cite{ref6}. It is not completely a volume impact since the $\chi^{2}$ best-fit replaces by $ \Delta \chi^{2}_{eff} = - 11$ contrasted with $\Lambda$CDM base model when including the one extra curvature parameter. The explanations behind the draw towards negative estimations of $\Omega_{k}$ are examined finally in \cite{ref5}. They are basically equivalent to those that lead to the inclination for $A_{L} > 1$, albeit marginally exacerbated for the situation of curvature since the low multipoles additionally fit the low- $ \ell$temperature probability marginally better if $\Omega_{k} < 0$. Similarly as with the $A_{L} > 1$ inclination, the joint Planck polarization result is not vigorous at the around $0.5\sigma$ level to displaying of the polarization
probabilities, with the CamSpec $ TT, TE, EE + lowE$ probability giving $\Omega_{k} = - 0.037^{+0.019}_{-0.014}$.\\

Closed models anticipate generously higher lensing amplitudes than in $L$CDM, so consolidating with the lensing reconstruction
(which is steady with a flat model) pulls parameters back into consistency with a spatially flat universe to well inside 2$\sigma$:\\

\hspace{3cm} $\Omega_{k} = - 0.01064\pm0.0065$ \qquad ( TT, TE, EE + lowE
+ lensing, 68 \%)\\

The limitation can be additionally honed by joining the Planck information with BAO information; this convincingly breaks the geometric decadence to give\\

\hspace{3cm} $\Omega_{k} = 0.0007\pm 0.0019$ \qquad ( TT, TE, EE + lowE +lensing + BAO, 68 \%)\\

The joint outcomes propose our Universe is spatially flat to a 1$\sigma$
the exactness of 0.2 \% \cite{ref6}.

\section{ Cosmological behaviour of the  interacting THDE model}
Substituting  Eq. (\ref{eq12}) in Eq. (\ref{eq8}), and combining with Eq. (\ref{eq11}), we obtained
\begin{eqnarray}
\label{eq15}
\omega _D= -\frac{\left(\Omega _K+1\right) \left(b^2 \left(\Omega _K+1\right)+(\delta -1) \Omega _D\right)}{\Omega _D \left((\delta -2) \Omega _D+\Omega _K+1\right)},
\end{eqnarray}

and
\begin{eqnarray}
\label{eq16}
q = -\frac{\left(\Omega _K+1\right) \left(\left(3 b^2-1\right) \left(\Omega _K+1\right)+(2 \delta -1) \Omega _D\right)}{2 \left((\delta -2) \Omega _D+\Omega _K+1\right)}.
\end{eqnarray}

For describing cosmic evolution, we have obtained cosmological parameters, deceleration parameter $q$ and THDE EoS parameter $\omega_{D}$ which are given by Eqs. (\ref{eq15}) and (\ref{eq16}). In Fig. 1, we present the behaviour of $q$
against redshift (z) for distinct Tsallis model parameters $\delta$ and $b^{2}$ and also different spatial curvature contributions of the universe. In the first column
of Fig. 1, the interaction term is absent in THDE model ($b^{2} = 0.0$). In the second column, we have taken interaction
($b^{2} =0.04$) and the third column, the
interaction term is considered as ($b^{2} =0.16$). The progress from decelerated phase $(q > 0)$ to accelerated phase $(q < 0)$ takes place in all the columns of Fig. 1. Moreover, the difference between them is minor but our results show that the transition from deceleration to accelerated phase is fully consistent with the observational data.\\

\begin{figure}
	\begin{center}
		\includegraphics[width=5.5cm,height=6.5cm, angle=0]{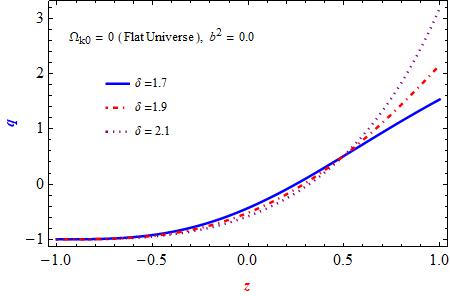}
		\includegraphics[width=5.5cm,height=6.5cm, angle=0]{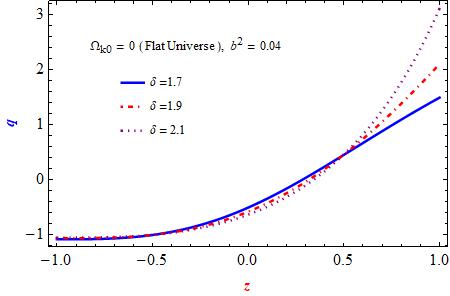}
		\includegraphics[width=5.5cm,height=6.5cm, angle=0]{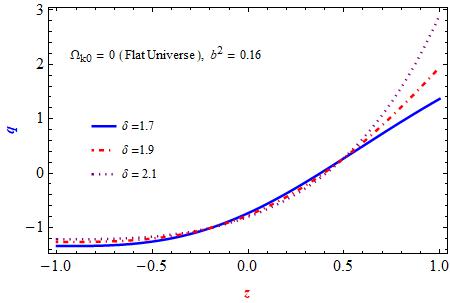}
		\includegraphics[width=5.5cm,height=6.5cm, angle=0]{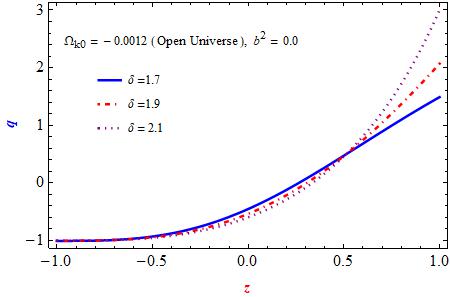}
		\includegraphics[width=5.5cm,height=6.5cm, angle=0]{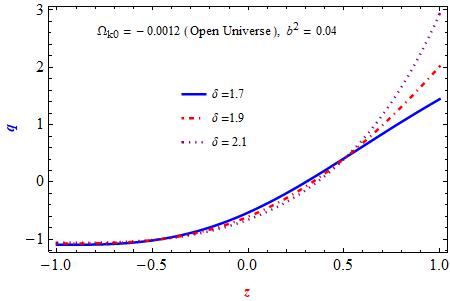}
		\includegraphics[width=5.5cm,height=6.5cm, angle=0]{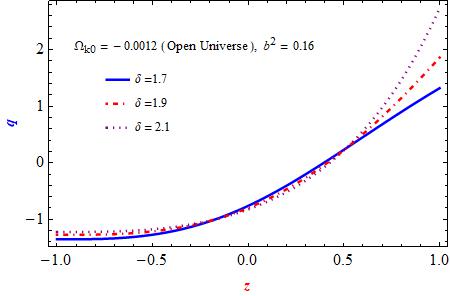}
		\includegraphics[width=5.5cm,height=6.5cm, angle=0]{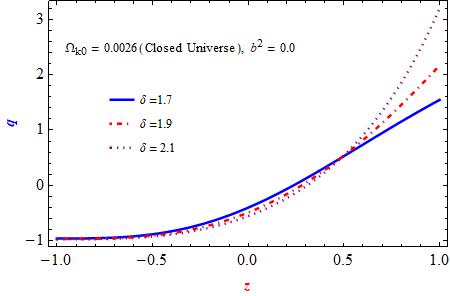}
		\includegraphics[width=5.5cm,height=6.5cm, angle=0]{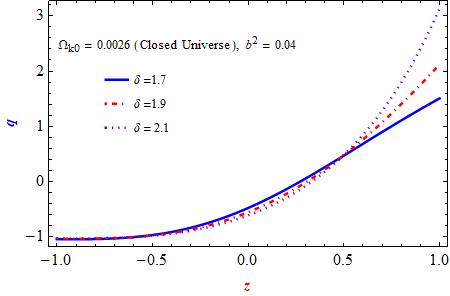}
		\includegraphics[width=5.5cm,height=6.5cm, angle=0]{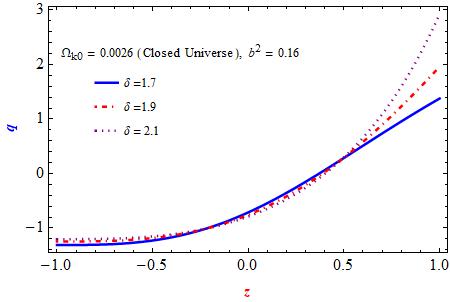}
		\caption { The evolutionary behaviour  of  deceleration parameter  ($q$)  in THDE model against redshift $z$ for different cases of Tsallis parameter $\delta = 1.7$, $\delta = 1.9$ and $\delta = 2.1$ and interaction term $b^{2}$. Selected graphs are plotted for $\Omega_{k0}$= $ 0$, $-0.0012$ and  $0.0026$  corresponding to   flat, open  and  colsed  universes, respectively, in the light of Planck 2018 results VI- LCDM base cosmology observational data.} 
		
	\end{center}
\end{figure}

The behaviour of THDE EoS parameter $\omega_{D}$ against redshift $z$ for
THDE with the apparent horizon cutoff has been plotted in Fig.
2, for different Tsallis model parameter $\delta$ and $b^{2}$ and also distinct spatial curvature contributions. In the first column of Fig. 2,
$\omega_{D}$ is graphed where the interaction term between DE and DM is absent ($b^{2} = 0.0$). It is clear from this column that for any spatial curvature,
the phantom divide line cannot be crossed and EoS parameter is quintessence like for $\delta =1.7, 1.9$ while phantom like for $\delta =2.1$. In the second and third columns, we have considered the interaction between dark energy and dark matter. For ($b^{2} = 0.04$), the phantom divide line is
reached for $\delta =1.7$ and various spatial curvature. In the third column, ($b^{2} = 0.16$), the behaviour of $\omega_{D}$ is not like the
previous situations and it is phantom like not crossing phantom divide line for all $\delta$ and contributions of spatial curvature.
 
 	 \begin{figure}
 	\begin{center}
 		\includegraphics[width=5.5cm,height=6.5cm, angle=0]{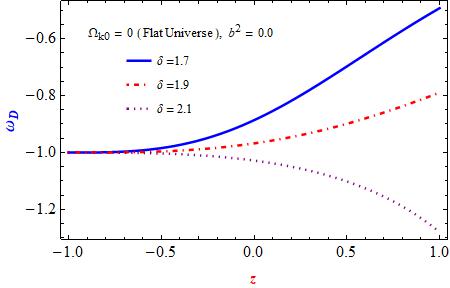}
  		\includegraphics[width=5.5cm,height=6.5cm, angle=0]{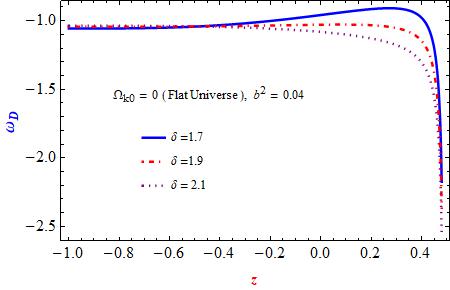}
  		\includegraphics[width=5.5cm,height=6.5cm, angle=0]{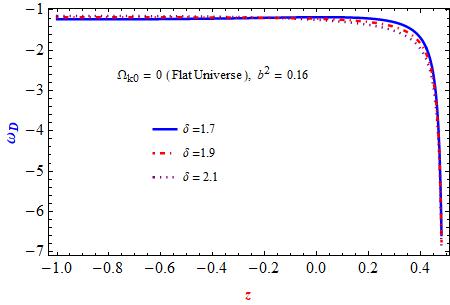}
  		\includegraphics[width=5.5cm,height=6.5cm, angle=0]{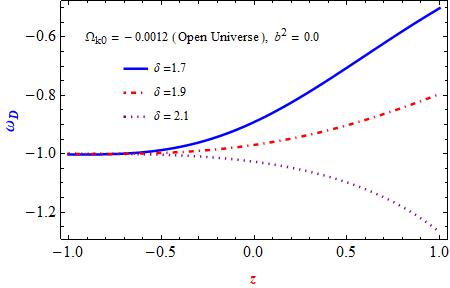}
  		\includegraphics[width=5.5cm,height=6.5cm, angle=0]{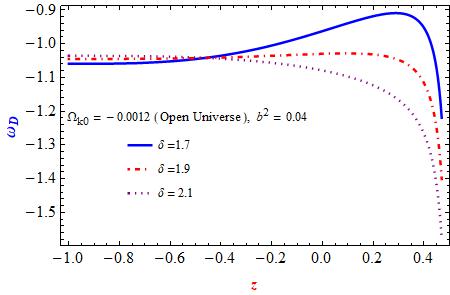}
  		\includegraphics[width=5.5cm,height=6.5cm, angle=0]{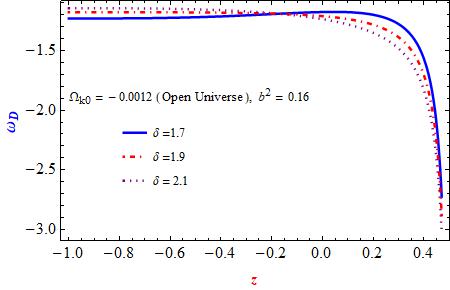}
  		\includegraphics[width=5.5cm,height=6.5cm, angle=0]{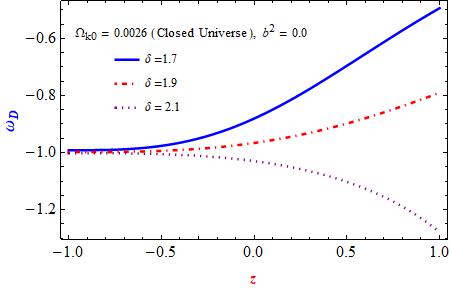}
  		\includegraphics[width=5.5cm,height=6.5cm, angle=0]{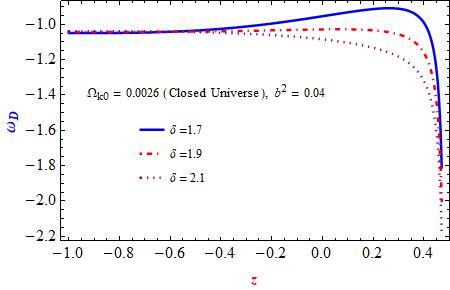}
  		\includegraphics[width=5.5cm,height=6.5cm, angle=0]{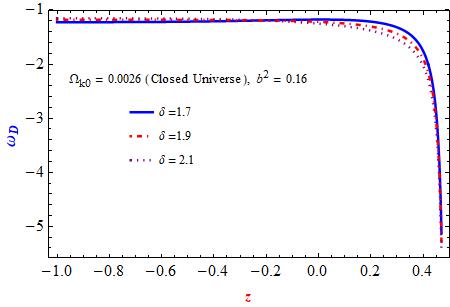}
  		\caption {The evolution of equation of state parameter  ($\omega_{D}$)  in THDE model against redshift $z$ for different cases of Tsallis parameter $\delta = 1.7$, $\delta = 1.9$ and $\delta = 2.1$ and interaction term $b^{2}$ . Selected graphs are plotted for $\Omega_{k0}$= $ 0$, $-0.0012$ and  $0.0026$  corresponding to   flat, open  and  closed  universes, respectively, in the light of Planck 2018 results VI- LCDM base cosmology observational data.} 
  		
  	\end{center}
 \end{figure}
  
  \begin{figure}
  	\begin{center}
 		\includegraphics[width=5.5cm,height=6.5cm, angle=0]{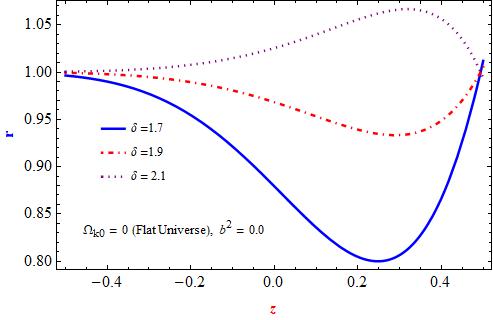}
 		\includegraphics[width=5.5cm,height=6.5cm, angle=0]{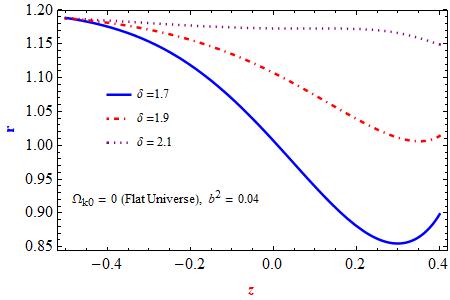}
  		\includegraphics[width=5.5cm,height=6.5cm, angle=0]{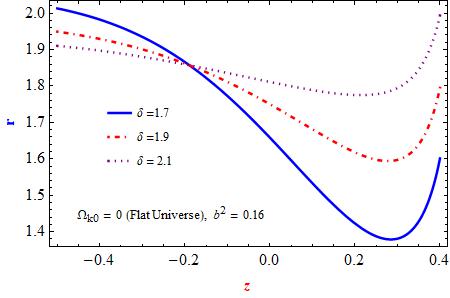}
  		\includegraphics[width=5.5cm,height=6.5cm, angle=0]{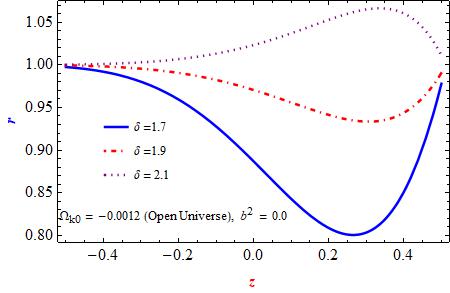}
  		\includegraphics[width=5.5cm,height=6.5cm, angle=0]{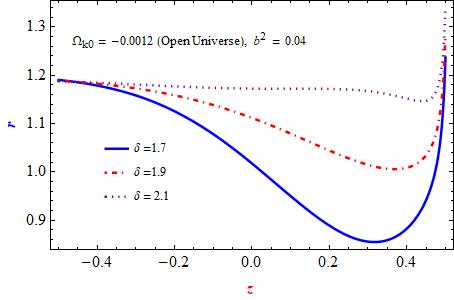}
  		\includegraphics[width=5.5cm,height=6.5cm, angle=0]{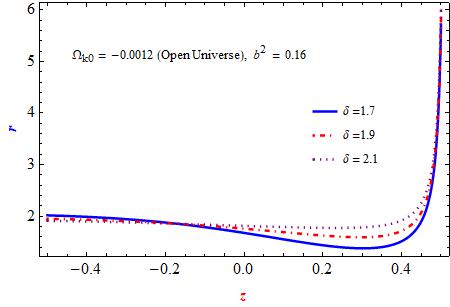}
  	\includegraphics[width=5.5cm,height=6.5cm, angle=0]{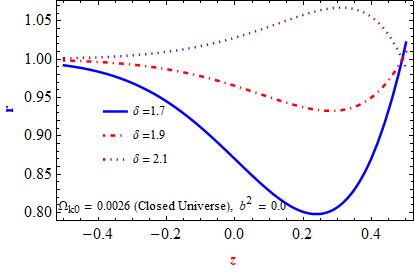}
  		\includegraphics[width=5.5cm,height=6.5cm, angle=0]{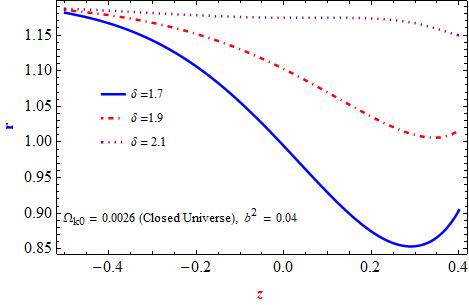}
  		\includegraphics[width=5.5cm,height=6.5cm, angle=0]{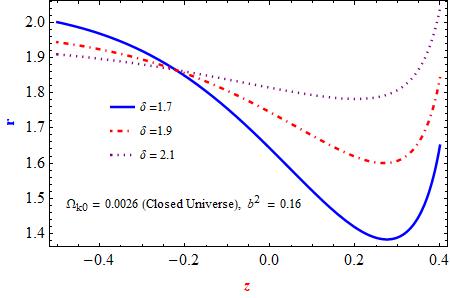}
  		\caption {The evolutionary behaviour  of first statefinder parameter  ($r$)  in THDE model against redshift $z$ for different cases of Tsallis parameter $\delta = 1.7$, $\delta = 1.9$ and $\delta = 2.1$ and interaction term $b^{2}$ . Selected graphs are plotted for $\Omega_{k0}$= $ 0$, $-0.0012$ and  $0.0026$  corresponding to   flat, open  and  closed  universes, respectively, in the light of Planck 2018 results VI- LCDM base cosmology observational data. } 
  		
  	\end{center}
  \end{figure}
  \begin{figure}
  	\begin{center}
  		\includegraphics[width=5.5cm,height=6.5cm, angle=0]{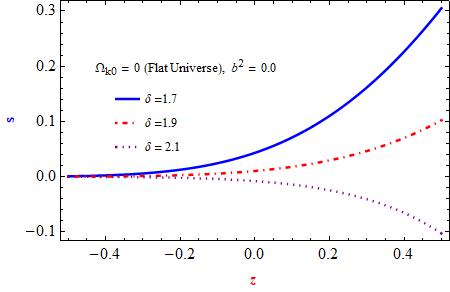}
 		\includegraphics[width=5.5cm,height=6.5cm, angle=0]{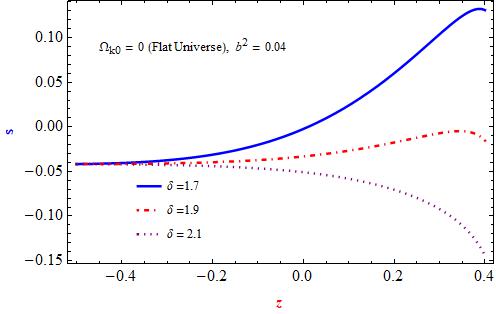}
  	\includegraphics[width=5.5cm,height=6.5cm, angle=0]{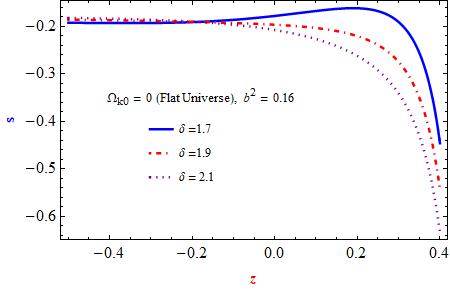}
  	\includegraphics[width=5.5cm,height=6.5cm, angle=0]{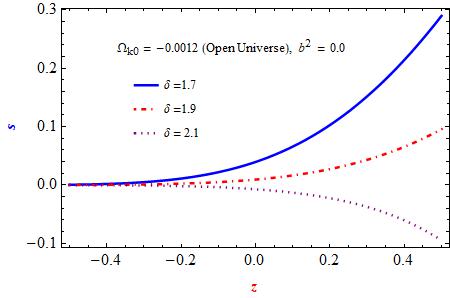}
  		\includegraphics[width=5.5cm,height=6.5cm, angle=0]{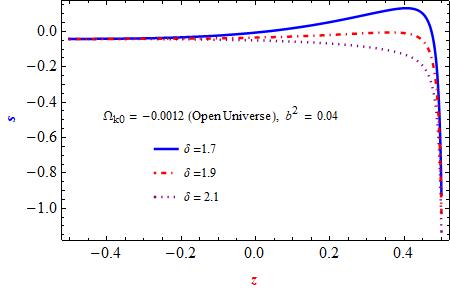}
  		\includegraphics[width=5.5cm,height=6.5cm, angle=0]{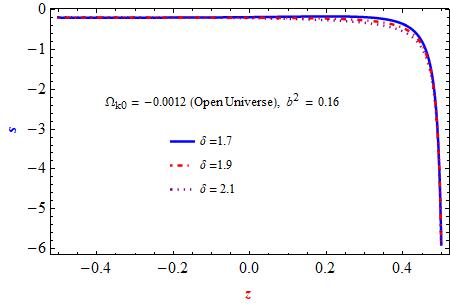}
  		\includegraphics[width=5.5cm,height=6.5cm, angle=0]{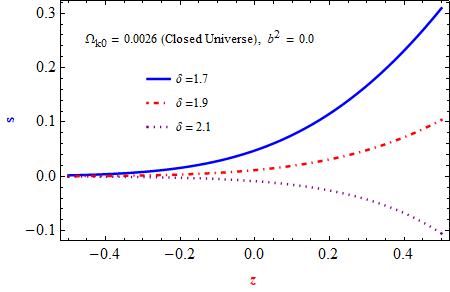}
  		\includegraphics[width=5.5cm,height=6.5cm, angle=0]{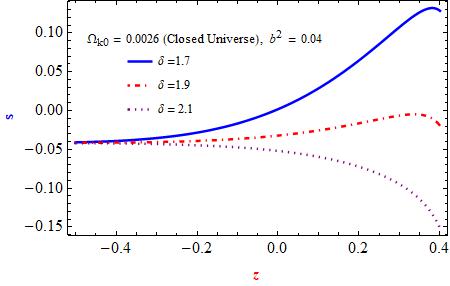}
  		\includegraphics[width=5.5cm,height=6.5cm, angle=0]{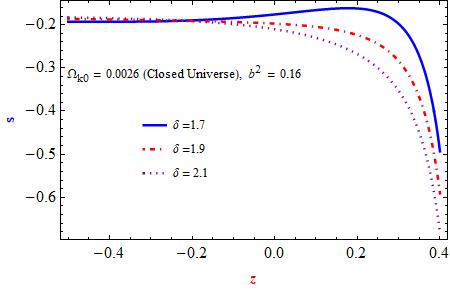}
  		\caption {The evolutionary behaviour  of the second statefinder parameter  ($s$)  in THDE model against redshift $z$ for different cases of Tsallis parameter $\delta = 1.7$, $\delta = 1.9$ and $\delta = 2.1$ and interaction term $b^{2}$ . Selected graphs are plotted for $\Omega_{k0}$= $ 0$, $-0.0012$ and  $0.0026$  corresponding to   flat, open  and  closed  universes, respectively, in the light of Planck 2018 results VI- LCDM base cosmology observational data.} 
  		
  	\end{center}
  \end{figure}

\section{Statefinder and $\omega_{D}-\omega_{D}'$ pair analysis for  interacting THDE}

Now, we discuss the statefinder analysis of the THDE
model. It ought to be referenced that the statefinder analysis with $\omega_{D}-\omega_{D}'$ pair for THDE in the flat universe examined in particular in \cite{ref48,ref49}, where the attention is put on the analysis of the various interims of parameter $\delta$. The statefinder diagnosis and $\omega_{D} - \omega_{D}^{'}$ plane diagnostic to the THDE (non - interacting ) model
in the non-flat universe with the apparent horizon as IR cutoff \cite {ref51}.
In \cite{ref48,ref49,ref51}, it has been exhibited that
from the statefinder perspective, $\delta$ and spatial curvature contribution assumes a key role. Here we need to concentrate on the statefinder analysis of the contribution of the interaction term. It is important that in the non-flat universe the $\Lambda$CDM model does not compare a fixed point in the statefinder plane, it displays an evolution trajectory as\\

\begin{eqnarray}
\Big(s,r\Big)_{nonflat-\Lambda CDM} = \Big(0,\Omega_{total}\Big)
\end{eqnarray}

The statefinder parameters $r$ and $s$ may also be expressed in terms of EoS parameter and energy density as follows \cite{ref38,ref40} :
\begin{eqnarray}
\label{eq20}
r= \Omega_{total}+ \frac{9}{2}\omega_{D}(1+\omega_{D})\Omega_{D}-\frac{3}{2}\omega_{D}^{'}\Omega_{D}
\end{eqnarray}
\begin{eqnarray}
\label{eq21}
s= 1+ \omega_{D}-\frac{1}{3}\frac{\omega_{D}^{'}}{\omega_{D}}
\end{eqnarray}
 here the parameter $r$
is also known as cosmic jerk and $ \Omega_{total}= 1+  \Omega_{K} =  \Omega_{m}+ \Omega_{D}$, is the total energy density.\\

\begin{figure}
	\begin{center}
		\includegraphics[width=5.5cm,height=6.5cm, angle=0]{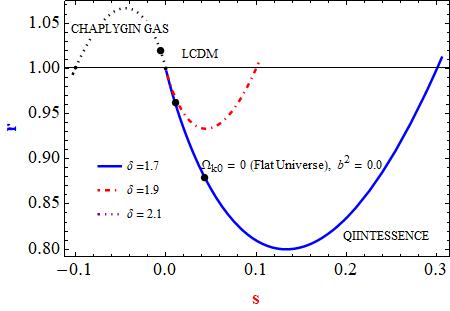}
		\includegraphics[width=5.5cm,height=6.5cm, angle=0]{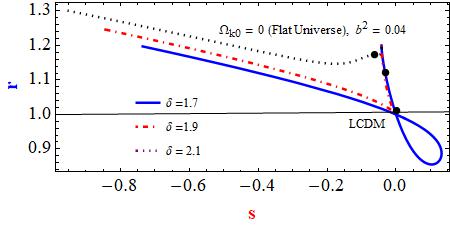}
		\includegraphics[width=5.5cm,height=6.5cm, angle=0]{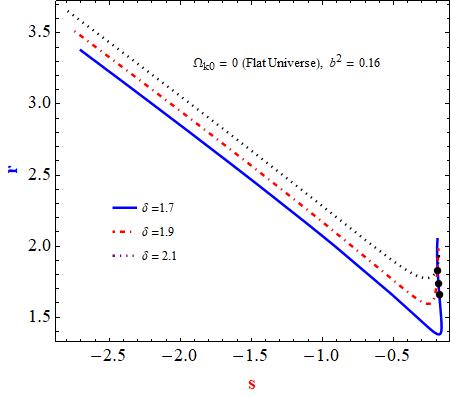}
		\includegraphics[width=5.5cm,height=6.5cm, angle=0]{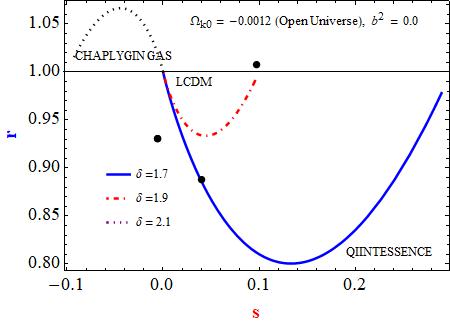}
		\includegraphics[width=5.5cm,height=6.5cm, angle=0]{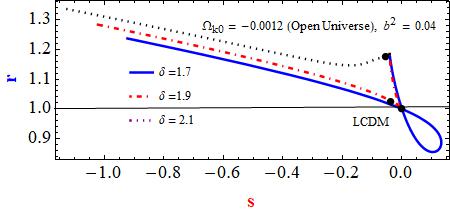}
		\includegraphics[width=5.5cm,height=6.5cm, angle=0]{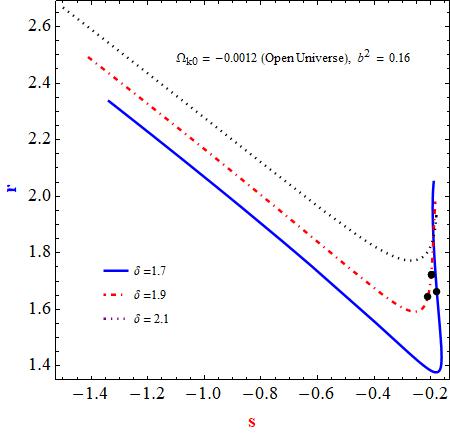}
		\includegraphics[width=5.5cm,height=6.5cm, angle=0]{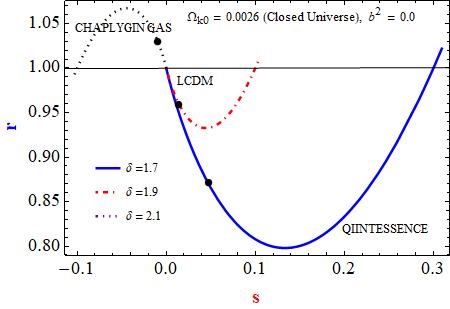}
		\includegraphics[width=5.5cm,height=6.5cm, angle=0]{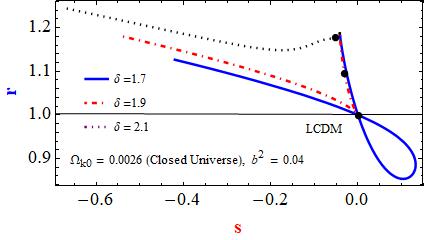}
		\includegraphics[width=5.5cm,height=6.5cm, angle=0]{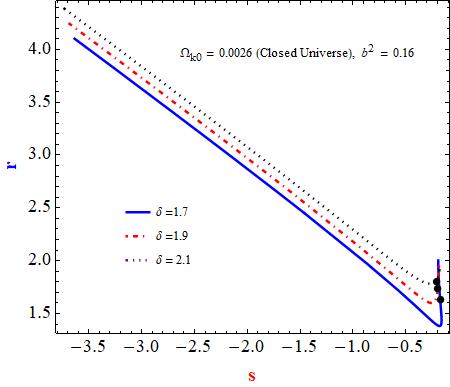}
		\caption {The  evolutionary trajectory in the  $s-r$ plane in THDE model for different cases of Tsallis parameter $\delta = 1.7$, $\delta = 1.9$ and $\delta = 2.1$ and interaction term $b^{2}$ . Selected graphs are plotted for $\Omega_{k0}$= $ 0$, $-0.0012$ and  $0.0026$  corresponding to   flat, open  and  closed  universes, respectively, in the light of Planck 2018 results VI- LCDM base cosmology observational data. LCDM corresponds the fixed point $(0, 1)$. The present values of ($s_{0}, r_{0}$) are represented by solid- dots circles.} 
		
	\end{center}
\end{figure}
\begin{figure}
	\begin{center}
		\includegraphics[width=5.5cm,height=6.5cm, angle=0]{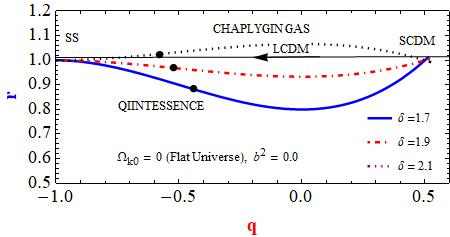}
		\includegraphics[width=5.5cm,height=6.5cm, angle=0]{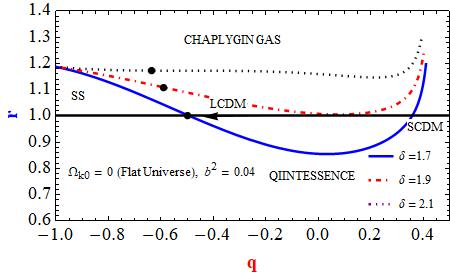}
		\includegraphics[width=5.5cm,height=6.5cm, angle=0]{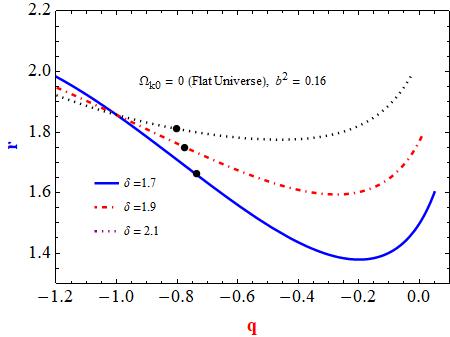}
		\includegraphics[width=5.5cm,height=6.5cm, angle=0]{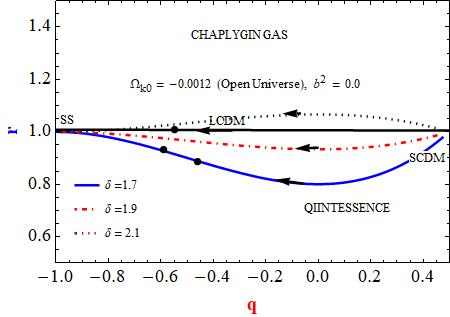}
		\includegraphics[width=5.5cm,height=6.5cm, angle=0]{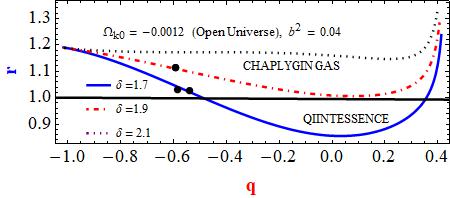}
		\includegraphics[width=5.5cm,height=6.5cm, angle=0]{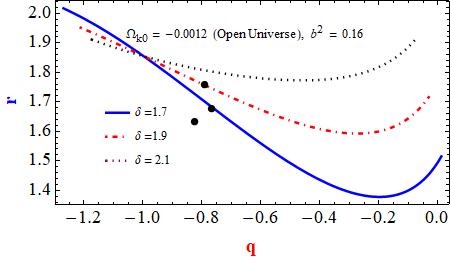}
		\includegraphics[width=5.5cm,height=6.5cm, angle=0]{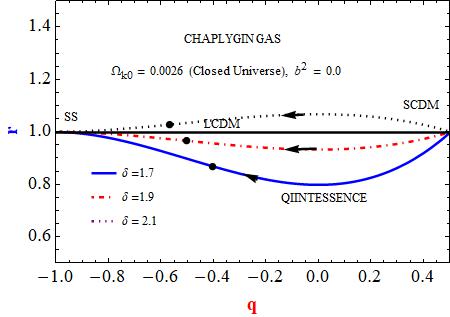}
		\includegraphics[width=5.5cm,height=6.5cm, angle=0]{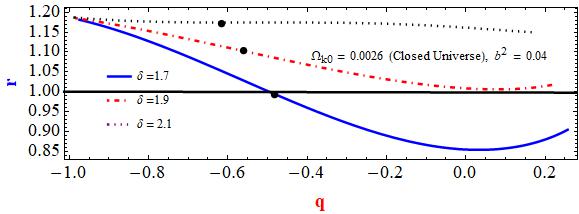}
		\includegraphics[width=5.5cm,height=6.5cm, angle=0]{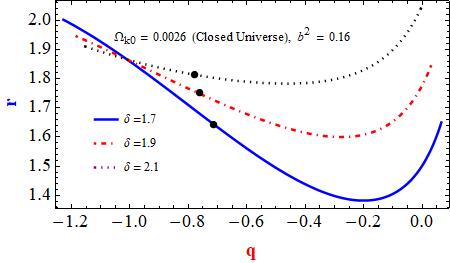}
		\caption {The evolutionary trajectory in the  $q-r$ plane in THDE model for different cases of Tsallis parameter $\delta = 1.7$, $\delta = 1.9$ and $\delta = 2.1$ and interaction term $b^{2}$ . Selected graphs are plotted for $\Omega_{k0}$= $ 0$, $-0.0012$ and  $0.0026$  corresponding to   flat, open  and  closed  universes, respectively, in the light of Planck 2018 results VI- LCDM base cosmology observational data. The de - sitter expansion - the steady state (SS) is the fixed point $(-1, 1)$, and $(0.5, 1)$ denotes the SCDM (matter dominated) universe. The present values of ($q_{0}, r_{0}$) are represented by solid- dots circles.} 
		
	\end{center}
\end{figure}
\begin{figure}
	\begin{center}
		\includegraphics[width=5.5cm,height=6.5cm, angle=0]{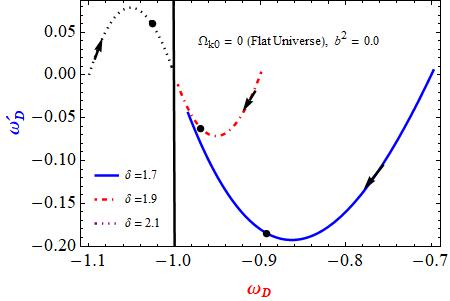}
		\includegraphics[width=5.5cm,height=6.5cm, angle=0]{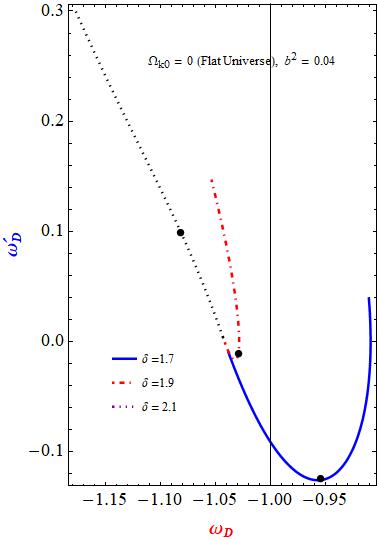}
		\includegraphics[width=5.5cm,height=6.5cm, angle=0]{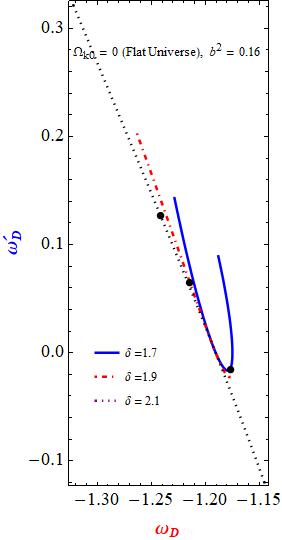}
		\includegraphics[width=5.5cm,height=6.5cm, angle=0]{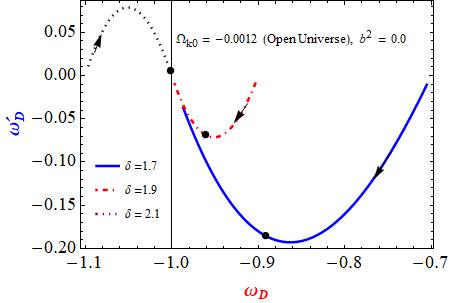}
		\includegraphics[width=5.5cm,height=6.5cm, angle=0]{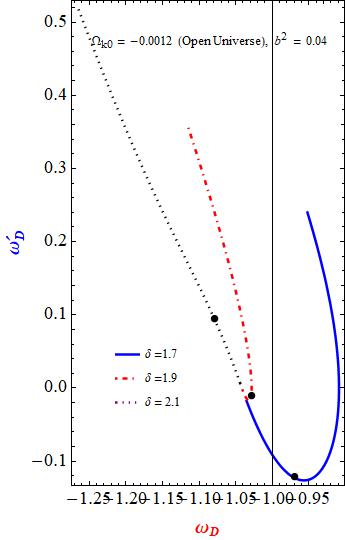}
		\includegraphics[width=5.5cm,height=6.5cm, angle=0]{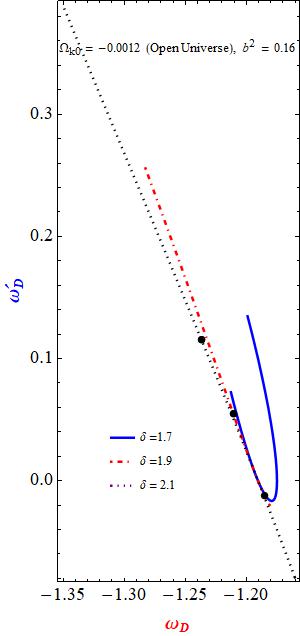}
		\includegraphics[width=5.5cm,height=6.5cm, angle=0]{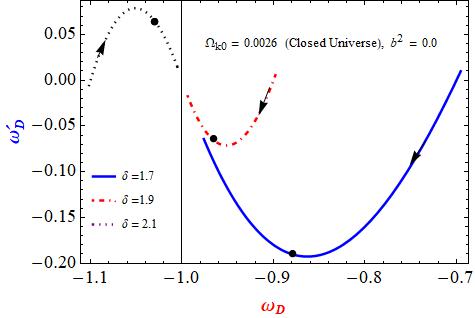}
		\includegraphics[width=5.5cm,height=6.5cm, angle=0]{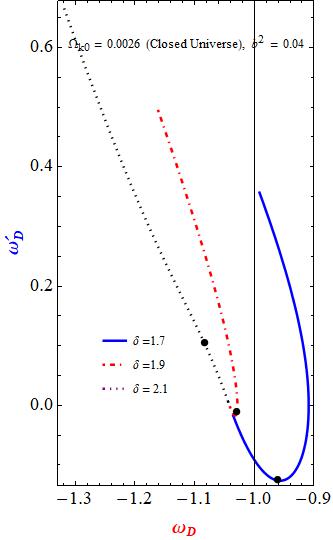}
		\includegraphics[width=5.5cm,height=6.5cm, angle=0]{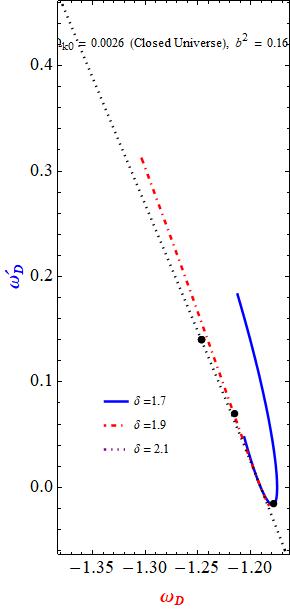}
		\caption {The THDE evolution trajectories in the $\omega_{D}- \omega_{D}^{'}$  plane in THDE model for different cases of Tsallis parameter $\delta = 1.7$, $\delta = 1.9$ and $\delta = 2.1$ and interaction term $b^{2}$ . Selected graphs are plotted for $\Omega_{k0}$= $ 0$, $-0.0012$ and  $0.0026$  corresponding to   flat, open  and  closed  universes, respectively, in the light of Planck 2018 results VI- LCDM base cosmology observational data. The present values of ($\omega_{D}- \omega_{D}^{'}$) are represented by solid- dots circles. } 
		
	\end{center}
\end{figure}

 The statefinder parameter are obtained for THDE model as :
\begin{eqnarray}
\label{eq23}
 r&=& \frac{9 \left(\Omega _K+1\right) \left(b^2 \left(\Omega _K+1\right)+(\delta -1) \Omega _D\right) \left(b^2 \left(\Omega _K+1\right){}^2-(\delta -2) \Omega _D^2+(\delta -2) \Omega _D \left(\Omega _K+1\right)\right)}{2 \Omega _D \left((\delta -2) \Omega _D+\Omega _K+1\right){}^2} \nonumber \\
&+&  
\frac{9 (\delta -1) \left(\Omega _K+1\right) \left(\left(b^2-1\right) \left(\Omega _K+1\right)+\Omega _D\right) \left(2 b^2 (\delta -2) \Omega _D \left(\Omega _K+1\right)+b^2 \left(\Omega _K+1\right){}^2+(\delta -2) (\delta -1) \Omega _D^2\right)}{2 \left((\delta -2) \Omega _D+\Omega _K+1\right){}^3}\nonumber\\
&+&1
\end{eqnarray}

\begin{eqnarray}
\label{eq24}
s& =&  -\frac{\left(\Omega _K+1\right) \left(b^2 \left(\Omega _K+1\right)+(\delta -1) \Omega _D\right)}{\Omega _D \left((\delta -2) \Omega _D+\Omega _K+1\right)} \nonumber \\
&-& \frac{(\delta -1) \left(\left(b^2-1\right) \left(\Omega _K+1\right)+\Omega _D\right) \left(2 b^2 (\delta -2) \Omega _D \left(\Omega _K+1\right)+b^2 \left(\Omega _K+1\right){}^2+(\delta -2) (\delta -1) \Omega _D^2\right)}{\left((\delta -2) \Omega _D+\Omega _K+1\right){}^2 \left(b^2 \left(\Omega _K+1\right)+(\delta -1) \Omega _D\right)}\nonumber\\
&+&1
\end{eqnarray}

\begin{landscape}
\setlength{\arrayrulewidth}{0.3mm}
\setlength{\tabcolsep}{0.5pt}
\renewcommand{\arraystretch}{1.9}
	\begin{table}
	\caption{The present values of the parameters r, s, q, $\omega_{D}$ and $\omega_{D}^{'}$ for different interaction term $b^2$ and  Tsallis parameter $\delta$  in different spatial curvature contributions.}
	\begin{center}
	---------------------
	
		\begin{tabular}{|c|c|c|c|c|c|c|c|c|c|c|}
			
			\hline
			Curvature contribution & Interaction Term  & \multicolumn{3}{|c|}{( Flat Universe) $\Omega_{k}$ = 0.00}  & \multicolumn{3}{|c|}{ (Closed Universe) $\Omega_{k}$ = 0.0026}&\multicolumn{3}{|c|}{( Open Universe) $\Omega_{k}$ = - 0.0012}  \\
			\hline
			\small Parameter& $b^2$ &\small $\delta$ = 1.7&\small $\delta$ = 1.9&\small $\delta$ = 2.1&\small $\delta$ = 1.7&\small $\delta$ = 1.9&\small $\delta$ = 2.1&\small $\delta$ = 1.7& \small$\delta$ = 1.9&\small $\delta$ = 2.1\\
			\hline
			\multirow{3}{*}{r} & 0 & 0.879249 & 0.968279 & 1.02546 & 0.871891 & 0.96595 & 1.02756 & 0.879283& 1.00978 & 0.930115\\ \cline{2-11}
			
			& 0.04 & 1.00683 & 1.10732	 & 1.1729 & 0.996229 & 1.10267 & 1.17326 & 1.01169& 1.11977 & 1.02612\\
			\cline{2-11}

			& 0.16 & 1.65848 & 1.75028 & 1.81149 & 1.64072 & 1.7417 & 1.80981 & 1.67383& 1.72298 & 1.63586\\
			\cline{2-11}
			\hline
			\multirow{3}{*}{s} & 0 & 0.0432623 & 0.0104058 & -0.00786095 & 0.0469951 & 0.0113922 & -0.00865601 &0.0402229 & 0.00961128 & -0.00722562\\ \cline{2-11}
			
			& 0.04 & -0.00226172 & -0.0331028 & -0.0507562 & 0.00127688 & -0.0322588 & -0.05167 & -0.00514395& -0.0337841 & -0.0500265\\ \cline{2-11}
			
			& 0.16 & -0.177847 & -0.196276 & -0.207479 & -0.176272 & -0.197057 & -0.209802 & -0.179154& -0.195662 & -0.205626\\
			\cline{2-11}
			\hline
			\multirow{3}{*}{q} & 0 & -0.43038& -0.516129& -0.579439 & -0.40737 & -0.494998 & -0.560136 & -0.450213& -0.534327 & -0.596076\\ \cline{2-11}
			
			& 0.04 & -0.506329 & -0.580645 & -0.635514 & -0.483103 & -0.5595582 & -0.616432 & -0.52642& -0.598849 & -0.652021\\ \cline{2-11}
			
			& 0.16 & -0.734177 & -0.774194 & -0.803738 & -0.710303 &-0.753334 & -0.785321 & -0.755041& -0.792416 & -0.819854\\
			\cline{2-11}
			\hline
			\multirow{3}{*}{$\omega_{D}$} & 0 & -0.886076 & -0.967742 & -1.02804 & -0.881263 & -0.966248 & -1.02942 & -0.890151& -0.968998 & -1.02688\\ \cline{2-11}
			
			& 0.04 & -0.958409 & -1.02919 & -1.08144 & -0.954712 & -1.02888 & -1.08402 & -0.961586& -1.02948 & -1.07932\\ \cline{2-11}
			
			& 0.16 & -1.17541 & -1.21352 & -1.24166 & -1.17506 & -1.21679 & -1.24781 & -1.17589& -1.21093 & -1.23665\\
			\cline{2-11}
			\hline
			\multirow{3}{*}{$\omega_{D}^{'}$} & 0 & -0.187835 & -0.063442 & 0.0622264 & -0.189671 & -0.0648163 & 0.0641281 & -0.185934& -0.0621829 & 0.0605497\\ \cline{2-11}
			
			& 0.04 & -0.126087 & -0.0120938& 0.0995545 & -0.126055 & -0.0104181 & 0.105202 & -0.125654& -0.0132928 & 0.0948574\\ \cline{2-11}
			
			& 0.16 & -0.00860378 & 0.0627706 & 0.127308 & -0.0042782 &0.0720401 & 0.142296 & -0.0115085& 0.0554534 & 0.11508\\
			\cline{2-11}

	\hline
		\end{tabular}
	\end{center}
\end{table}	
\end{landscape}
The first and second statefinder parameters against redshift have been plotted in Fig. 3 and 4. Fig. 3, presents the behaviour of first parameter $r$
against redshift (z) for distinct Tsallis model parameters $\delta$ and $b^{2}$ and also different spatial curvature contributions of the universe and Fig. 4, shows the behaviour of the second parameter $s$
against redshift (z) for distinct Tsallis model parameters $\delta$ and $b^{2}$ and also different spatial curvature contributions of the universe. In the first column
of Fig. 3 and Fig. 4, the interaction term is absent in THDE model ($b^{2} = 0.0$). In the second column, we have taken interaction
($b^{2} =0.04$) and the third column, the
interaction term is considered as ($b^{2} =0.16$). In the first column of both the figures, we see that, both, first and second statefinder parameter at low red-shift of THDE approaches that of $\Lambda$CDM i.e. $r=1$ and $s=0$, for different Tsallis model parameter $\delta$ and also open, flat and closed universes while, In the second and third column of both the figures, the behaviour of first and second parameter deviates significantly from the standard behaviour of $\Lambda$CDM.\\

We have plotted the evolution trajectories in the statefinder $(r, s)$ and $(r, q )$
planes for our THDE model in Fig. 5 and Fig. 6, for different Tsallis parameter $\delta$ and $b^{2}$ and also different contributions of the spatial curvature as
$\Omega_{K0}$ as 0, -0.0012 and 0.0026
corresponding to the flat, open and closed universes, respectively. From Fig. 5, in $(r, s)$ evolutionary plane, we see that the derived THDE model, for ($b^{2} = 0.0$), starts its evolutionary trajectories from different points for different curvature contribution and ends at $LCDM$ fixed point ($r=1, s=0$) in the future for $\delta=1.7, 1.9$ and for $\delta=2.1$, shows Chaplygin gas behaviour. In the second and third columns, it tends to be seen that the various curvature will prompt distinctive evolutionary behaviour in the statefinder plane.\\

From Fig. 6, in $(r, q)$ evolutionary plane, in the panel of first column, we observe that the Tsallis HDE model evolutionary trajectories starts from matter dominated universe i.e. SCDM ( $r = 1$, $q = 0.5$) in the past, and their evolutionary trajectories approaches the
point ($q = -1$, $r = 1$) in the future i.e. the de Sitter expansion ($SS$) for the open, flat and closed universes for $\delta=1.7, 1.9$ and for $\delta=2.1$, shows Chaplygin gas behaviour without interaction ($b^{2} = 0.0$). In the panel of the second and third column, we can see that, from Fig. 6, the distance from the de Sitter expansion increases as interaction increases. Setare et al. \cite{ref38}, have a diagnosis the HDE with future event horizon as IR cutoff in the non-flat universe where the focus was different values of parameter $c$ and the spatial curvature contributions.\\

Finally, we do the $\omega_{D}- \omega_{D}^{'}$ diagnostic for interacting non flat THDE model. In
this dynamical analysis, the fixed point $\omega_{D} = -1$, $\omega_{D}^{'}=0$ corresponds to the standard $LCDM$ in the $\omega_{D}-\omega_{D}^{'}$ plane. The evolutionary trajectories of $\omega_{D}^{'}$ and $\omega_{D}$ plane are shown in Fig. 7, for different Tsallis model parameter $\delta$ and also the different spatial curvature contributions. The first column of Fig. 7, where the
evolutionary trajectories without interaction term $b^{2} =0$ are plotted for THDE model. In this case, the trajectories approach to the point ($\omega_{D} = -1$, $\omega_{D}^{'} = 0$). In second
column panels, the trajectories are plotted for $b^{2} = 0,04$. In this case, $\omega_{D}$ crosses $-1$ for $\delta = 1.7$ while in the first case, it was not. In
third column panels, the evolutionary trajectories are shown for $b^{2}=0,16$. In this case, the trajectories are different from the previous cases. The EoS $\omega_{D}$ shows phantom behaviour.

\section{Conclusion}

 The interacting THDE model in the nonflat universe has been expanded considering apparent horizon as IR cutoff from the statefinder and $\omega_{D} - \omega_{D}'$ pair viewpoint in this paper. This expansion can be outlined as \\

We considered the evolution of the deceleration parameter, $q$, and EoS parameter, $\omega_{D}$, for different Tsallis parameter $\delta$ and $b^{2}$ and also distinct spatial curvature contributions corresponding to the flat, open and closed universes, respectively. We demonstrated that for any spatial curvature, in absence of curvature term, EoS cannot cross the $\omega_{D} =-1$ i.e. phantom divide line for the derived THDE model. To accomplish the phantom line at any curvature contribution, there is a requirement of the interaction term. The change from decelerated stage to quickened stage is reliant on parameters $\delta$ and $b^{2}$ just as the kind of spatial curvature of the cosmos. \\

We investigated the statefinder demonstrative and $\omega_{D}-\omega_{D}^{'}$ investigation for the non-flat universe of THDE model considering the interaction between dark matter and DE. The geometrical statefinder demonstrative and $\omega_{D}-\omega_{D}^{'}$ examination are helpful techniques to separate the different models of DE. Also, despite the fact that we are missing with respect to the standard theory for the DE, this theory is attempted to have a few highlights of a quantum gravity theory, which may be examined theoretically by considering the holographic rule of quantum gravity theory. So THDE model gives us an endeavour to investigate the nature of DE inside a structure of the fundamental theory. $\omega_{D} - \omega_{D}'$ pair and the statefinder pair plots demonstrate that the spatial curvature contributions and interaction effect in the model can be analyzed expressly in this technique. \\

The present values of the paramters $r, s, q, \omega_{D}$ and $\omega_{D}^{'}$ for different Tsallis parameter $\delta$ in different spatial curvature contributions and also for different interaction $b^{2}$ are summarised in Table 1.\\

\section*{Acknowledgments}
The authors are appreciative to the help of IUCAA, Pune, India, where a part of this work was completed.


\end{document}